# Hysteresis in the transfer characteristics of MoS$_2$ transistors

Antonio Di Bartolomeo[1,2], Luca Genovese[1], Filippo Giubileo[2], Laura Iemmo[1], Giuseppe Luongo[1,2], Tobias Foller[3] and Marika Schleberger[3]

[1]*Physics Department, University of Salerno, 84084 Fisciano, Salerno, Italy. Corresponding author email: adibartolomeo@unisa.it*

[2]*CNR-SPIN Salerno, 84084 Fisciano, Italy*

[3]*Fakultät für Physik and CENIDE, Universität Duisburg-Essen, Lotharstrasse 1, D-47057, Duisburg, Germany*

**Abstract**

We investigate the origin of the hysteresis observed in the transfer characteristics of back-gated field-effect transistors with an exfoliated MoS$_2$ channel. We find that the hysteresis is strongly enhanced by increasing either gate voltage, pressure, temperature or light intensity. Our measurements reveal a step-like behavior of the hysteresis around room temperature, which we explain as water-facilitated charge trapping at the MoS$_2$/SiO$_2$ interface. We conclude that intrinsic defects in MoS$_2$, such as S vacancies, which result in effective positive charge trapping, play an important role, besides H$_2$O and O$_2$ adsorbates on the unpassivated device surface. We show that the bistability associated to the hysteresis can be exploited in memory devices.

**Introduction**

Molybdenum disulfide (MoS$_2$) has recently become one of the most popular semiconductors from the family of the transition metal dichalcogenides [1]. Its bulk form consists of a stacking of two-dimensional layers, which are weakly bounded by van-der-Waals interactions. Each MoS$_2$ layer is actually formed by three atomic planes, with the Mo plane sandwiched between two S planes. Similar to graphene, the weak interlayer bonding enables monolayer MoS$_2$ cleavage, either by scotch-tape or liquid-phase exfoliation [2]. For larger scale production, chemical vapor deposition (CVD) from heated S and MoO$_3$ powders has become the standard production method [3]. The MoS$_2$ bandgap can be controlled by the number of layers: Bulk MoS$_2$ has an indirect bandgap of 1.2 eV while monolayer MoS$_2$ has a direct bandgap of 1.8 eV [4]. The large bandgap, combined with mechanical flexibility, makes MoS$_2$ suitable as channel in field effect transistors (FETs) for logic applications. Indeed, MoS$_2$ devices on SiO$_2$/Si substrates with On/Off current ratio up to $10^8$ have been reported [5-8]. However, the carrier mobility achieved in these devices remains limited to values more than a factor 10 below the theoretical phonon-limited value of 410 $cm^2V^{-1}s^{-1}$ on SiO$_2$ [9, 10], even when the mobility is enhanced by suppressed Coulomb scattering via dielectric screening in high-k materials deposited on top of the MoS$_2$ [11-13]. The low mobility has been mainly ascribed to short range scattering caused by structural defects, such as S vacancies and grain boundaries, or other forms of disorder related to the substrate, such as Coulomb traps, remote polar phonons and surface corrugations [14, 15].

The direct bandgap favors light absorption/emission and renders $MoS_2$ suitable for optoelectronic applications [16] in photodetectors [17-18], light emitters [19] or photovoltaic cells [20, 21]. In particular, $MoS_2$ phototransistors have shown persistent photoconductivity, whose origin has been attributed both to intrinsic or extrinsic causes. The intrinsic effect consists in photocharge trapping in localized band-tail states of $MoS_2$ [22, 23], while extrinsic causes include photocharge trapping in residues and adsorbates deposited on the top or bottom $MoS_2$ surface. Trap states can be created by dangling bonds in the underlying dielectric [23, 24], while top trap states are mainly caused by adsorbed moisture and oxygen molecules [25] or processing residues. A phenomenon related to the persistent photoconductivity, which is also related to charge transfer/trapping, is the hysteresis observed in the transfer characteristics of $MoS_2$ transistors. Hysteresis appears when gate voltage sweeps result in right or left shift of the transfer characteristic, with subsequent change of the transistor threshold voltage. Such phenomenon has been reported in several field effect devices based on nanostructured materials, such as carbon nanotube networks [26], single nanotubes [27] or graphene [28-30]. Hysteresis is in general an unwanted effect, which makes the transistor parameters dependent on the gate voltage sweep range, direction and time, and on the loading history. Therefore, it should be minimized or eliminated to improve the device stability and reliability. Nevertheless, hysteresis can be conveniently exploited to fabricate memory devices [28, 31-33].

The strong dependence on the gate voltage suggests that the hysteresis originates from charge redistribution under the gate electric stress, which can result in charge transfer, charge trapping or charge polarization. Despite the agreement on this general framework, the details of the mentioned mechanisms are still under debate. Some groups have suggested that charge traps are located at the interface between the $MoS_2$ and the $SiO_2$ [23, 34-36], other studies have shown that the adsorption/desorption of gases on the exposed $MoS_2$ surface by high vacuum or temperature annealing strongly affects the trapping process [37-42], while a recent work [43], besides the extrinsic factors, identifies the intrinsic charge transfer and trapping in sulfur vacancies or in other structural defects of the $MoS_2$ [44, 45] as the dominant cause of hysteresis.

In this paper, we fabricate back-gated, exfoliated-$MoS_2$ transistors and we investigate the effects of gate voltage stress and environmental conditions. Our study confirms that hysteresis is enhanced by $O_2$ or $H_2O$ molecules adsorbed on the device surface and can be quenched by reducing the pressure or the temperature. Our experimental findings suggest that hysteresis is strongly related to water, which facilitates charge transfer and trapping, and is favored when thermally-generated or photogenerated minority carriers (holes) become available in the n-type $MoS_2$ transistor. We suggest that intrinsic defects such as S vacancies or other interfacial states, which result in effective positive charge trapping, are an important cause of hysteresis.

**Experimental methods**

The fabrication of the devices (Figure 1 (a)) started with the mechanical exfoliation of MoS₂ flakes on heavily p-doped Si substrate ($0.01 - 0.05\ \Omega\ cm$ resistivity), covered by thermal oxide with $t_{SiO_2} = 285\ nm$ thickness. Before placing the markers needed for subsequent lithographic steps, the silicon substrate was cleaned with acetone and isopropanol using ultrasonic cleaner for 3 and 10 min, respectively. Monolayer MoS₂ flakes were selected by optical microscopy and confirmed by micro-Raman spectroscopy (Renishaw inVia, 532 nm wavelength). Figure 1(b) shows the $E_{2g}^1$ and $A_{1g}$ Raman modes, with $\sim 19.5\ cm^{-1}$ peak separation, which is the typical signature of monolayer MoS₂ [46]. Contact leads were patterned in a four-probe configuration by electron beam lithography and standard lift-off of evaporated Ni(5 nm)/Au(50 nm) films. The four leads enable multiple source/drain choices and four-probe measurements to eliminate any effect of the contact resistance [47]. No annealing was applied after the lift-off process and the device top surface remained air-exposed.

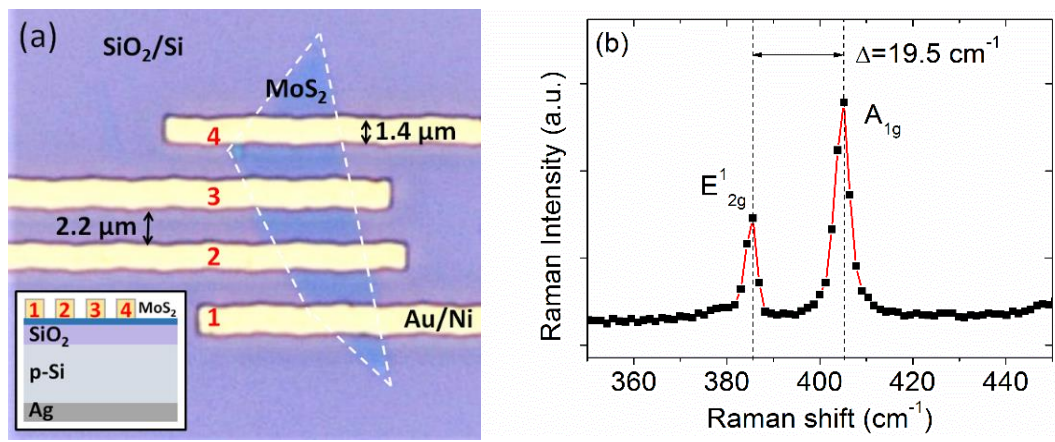

Figure 1. (a) Optical image of a monolayer MoS₂ flake (highlighted by dashed white lines) contacted with Ni/Au leads; the inset shows the schematic cross-section of the back-gated FET. (b) Raman spectrum of the MoS₂ flake.

Since low contact resistance ($\leq 1\ k\Omega\ \mu m$) is achieved with evaporated Ni on MoS₂ [13, 47], for transistor characterization, we typically adopted the two-probe configuration in voltage bias or sweeping mode. For the measurements discussed in the following, we used the innermost leads, labelled as 2 and 3 in Figure 1(a), as the source and the drain, and the heavily doped substrate as the gate of the transistor. Electrical measurements were performed in a Janis ST-500 cryogenic probe station connected to a Keithley 4200-SCS semiconductor parameter analyzer. If not otherwise stated, measurements were performed at atmospheric pressure, room temperature and in darkness.

**Results and discussion**

The linear current-voltage output characteristics, $I_{ds} - V_{ds}$, which are plotted in Figure 2(a) for several gate biases, $V_{gs}$, indicate ohmic contacts and transport in the near equilibrium regime (low $V_{ds}$). The channel resistance is clearly gate modulated. The effect of the gate is fully characterized by the $I_{ds} - V_{gs}$ transfer characteristic of Figure 2(b), where the current at the drain bias $V_{ds}$=30 mV is displayed both in linear and logarithmic scale. The FET behaves as an n-type MOS in depletion mode (normally-on transistor), indicating that the MoS$_2$ channel is n-doped. Native n-doping is commonly attributed to the omnipresent electron-donating sulfur vacancies especially present in exfoliated MoS$_2$ [15, 48-50], even though the n-doping character of sulfur vacancies has been disputed [51]. Alternatively, Re impurities, often present in natural MoS$_2$, have been proposed as another important n-type dopant [51].

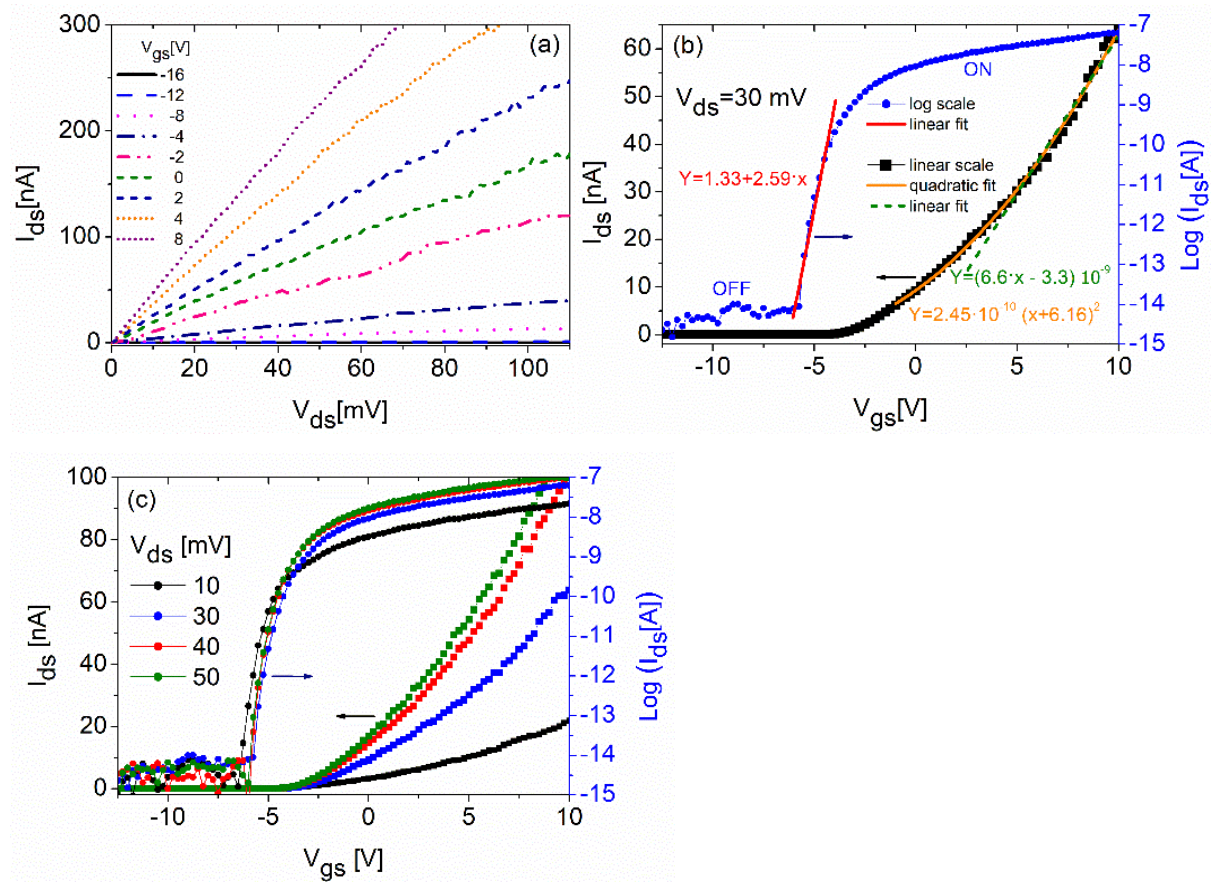

Figure 2. (a) $I_{ds} - V_{ds}$ output characteristics of MoS$_2$ back-gated FET. (b) $I_{ds} - V_{gs}$ transfer characteristic at $V_{ds}$ = 30 mV with current in logarithmic and linear scale, and linear and parabolic fitting curves. (c) $I_{ds} - V_{gs}$ transfer characteristics for various $V_{ds}$ with current in logarithmic and linear scale.

The polarity of a FET is determined by the type of charge carriers that can be injected from the source contact into the semiconductor channel. Accordingly, the widely reported n-type behavior for MoS$_2$ FETs [15, 52] has been interpreted as Fermi level pinning close to the conduction band edge of MoS$_2$, resulting from the metal-MoS$_2$ interaction. Such pinning favors electron injection through negligible Schottky barrier and suppresses hole transport by the formation of a large hole Schottky barrier [53, 54]. A p-type behavior can be achieved in MoS$_2$ FETs using high-workfunction metal contacts, such as MoO$_x$ with x<3 [54], or by exploiting substitutional acceptors such as niobium [48] in Nb/Pd contacts [55]. Figure 2 (a) and (b) support the hypothesis of alignment of the Fermi level near the conduction band edge of MoS$_2$, given the n-type behavior and the negligible Schottky barrier for electrons.

The current in logarithmic scale shows an On/Off ratio greater than $10^7$ and an exponential transition from the subthreshold to the conduction region corresponding to a subthreshold swing (which is the inverse of the slope of the straight line fitting the blue-circle curve in Figure 2 (b))

$$S = \left(\frac{dV_{gs}}{d(logI_{ds})}\right) \approx \frac{380 \ mV}{decade}.$$

The current in linear scale (black squares) has a quadratic dependence on $V_{gs}$ despite the device is at low $V_{ds}$, which is the triode region, where traditional FETs exhibit a linear dependence both on $V_{ds}$ and $V_{gs}$. This feature appears also at other $V_{ds}$ biases, as shown in Figure 2 (c), and is very common in literature [34, 37, 41], although rarely highlighted. The device behaves as it was in saturation, i.e. in the region where the current is limited by source injection. In this scenario, the quadratic $I_{ds} - V_{gs}$ curve is related to the electron injection rate from the contacts, which is affected by the gate. We can model such a feature by introducing a $V_{gs}$ dependent mobility: $\mu = \mu_V(V_{gs} - V_{th})$. Consequently, we express the FET channel current in the triode region as:

$$I_{ds} = \mu_V C_{ox} \frac{Z}{L} (V_{gs} - V_{th})^2 V_{ds}$$

where $\mu_V$ is expressed in $cm^2 V^{-2} s^{-1}$, $C_{ox} = \varepsilon_{SiO_2}/t_{SiO_2} = 1.21 \cdot 10^{-8} \ F cm^{-2}$ is the capacitance per unit area of the SiO$_2$ gate dielectric, $L = 2.2 \ \mu m$ and $Z = 6.4 \ \mu m$ are the channel length and width, respectively, and $V_{th}$ is the threshold voltage. From the quadratic fit shown in Figure 2 (b), we extract $V_{th} = -6.2 \ V$ and $\mu_V = 0.23 \ cm^2 V^{-2} s^{-1}$, which corresponds to a mobility $\mu = 3.7 \ cm^2 V^{-1} s^{-1}$ at $V_{gs} = 10 \ V$. We notice that the usual approach with the mobility evaluated as $\mu = \frac{1}{C_{ox}} \frac{L}{Z} \frac{1}{V_{ds}} \frac{dI_{ds}}{dV_{gs}}$, where $\frac{dI_{ds}}{dV_{gs}}$ is the slope of a linear fit of the $I_{ds} - V_{gs}$ curve (olive-dashed straight line in Figure 2 (b)), yields the slightly higher value $\mu = 6.2 \ cm^2 V^{-1} s^{-1}$. The mobilities obtained from the $I_{ds} - V_{gs}$ curves of Figure 2 (c) at different $V_{ds}$ have similar values. Mobilities in the range $0.05 - 50 \ cm^2 V^{-1} s^{-1}$ are common in FETs with MoS$_2$ monolayer channel on SiO$_2$ [38, 56-59] and are caused by intrinsic/extrinsic scattering centers and/or by Schottky barriers at the contacts [52, 60]. Based on the previous discussion, the Schottky barrier is negligible in our device and the low mobility is indicative of a high density of scatterers, such as intrinsic defects/traps or charged impurities. Since MoS$_2$ has a high surface-to-volume ratio, surface defects have a profound effect on charge transport. In this context, the high density of defects is favorable to the study of the hysteresis. The linear increase of the mobility with $V_{gs}$, that we have assumed, could be due to increased screening of Coulomb scatterers by an increasing carrier density as well as to the mentioned gate-dependent electron injection rate from the contacts. For weakly interacting metal contacts on MoS$_2$ layers, the contact resistance is limited by tunneling through the van der Waals gap and can be affected by the gate voltage [61]. Indeed, it has been demonstrated that the carrier mobility measured in a two terminal-configurations depends on the quality of the contacts [62].

Figure 3 (a) shows the transfer characteristics in a series of increasing $V_{gs}$ loops, and demonstrates that wider sweeping range, i.e. increased gate voltage stress, results in a wider hysteresis. This behavior is similar to what has been observed for instance on carbon nanotubes (CNTs) [26] or graphene [28] based transistors. The dominant mechanism of hysteresis there was identified as charge storage at the SiO$_2$ top surface below the CNT/graphene channel, through a mechanism involving Si-OH silanol groups and water molecules on the SiO$_2$ surface, which can act as electron trap centers [26, 63, 64]. In particular, H$_2$O molecules would mediate the transfer of electrons to and from the traps and/or provide extra traps by weakening some bonds in the top SiO$_2$ layers.

In Figure 3 (a), starting from zero, the voltage was swept first in the positive direction and then reversed towards negative values ("right-first" loop). For each successive curve the gate voltage range was increased by 8 V. Figures 3 (b) and (c), which plot only a part of the loop (the rising and falling branches, respectively) show that the negative voltage is more effective in left-shifting the transfer characteristic. This is further confirmed by Figures 3 (d)-(f), where the $V_{gs}$ loops is reversed (from zero to negative and then to positive voltages, i.e. "left-first loop"), which show a more pronounced left-shift on both the rising and falling branches. The hysteresis loop remains always confined to negative $V_{gs}$, which corresponds to an effective trapping/detrapping of positive charge. The trapped charge can be roughly estimated from the hysteresis width, $W = |V_{gs1}(I_w) - V_{gs2}(I_w)|$, defined as the absolute value of the difference between the two $V_{gs}$ values corresponding to the current $I_w = 10^{-12} A$ (Figure 3). Such a width (up to ~10 V in the wider loop) corresponds to a density of trapped charges $n_t = W \cdot \frac{C_{ox}}{|q|} \approx 8 \cdot 10^{11}$ cm$^{-2}$ (q is the electron charge). A possible source of positive charge are donor-like S mono-vacancies or more complex defects involving S vacancies, which act as positive trap centers. Indeed, an average defect density of ~$10^{13} cm^{-2}$, dominated by S vacancies, has been measured on exfoliated MoS$_2$ monolayer by aberration-corrected scanning transmission electron microscopy [45]. Then, the S vacancies alone would be more than enough to accumulate the charge needed to generate the observed hysteresis. The mechanism is illustrated by the band diagram of Figure 4, where S vacancies have been located at about 0.25-0.5 eV from the minimum of the conduction band [53, 65]. Being donor-like, these S vacancies states are neutral when occupied by electrons and positively charged when unoccupied.

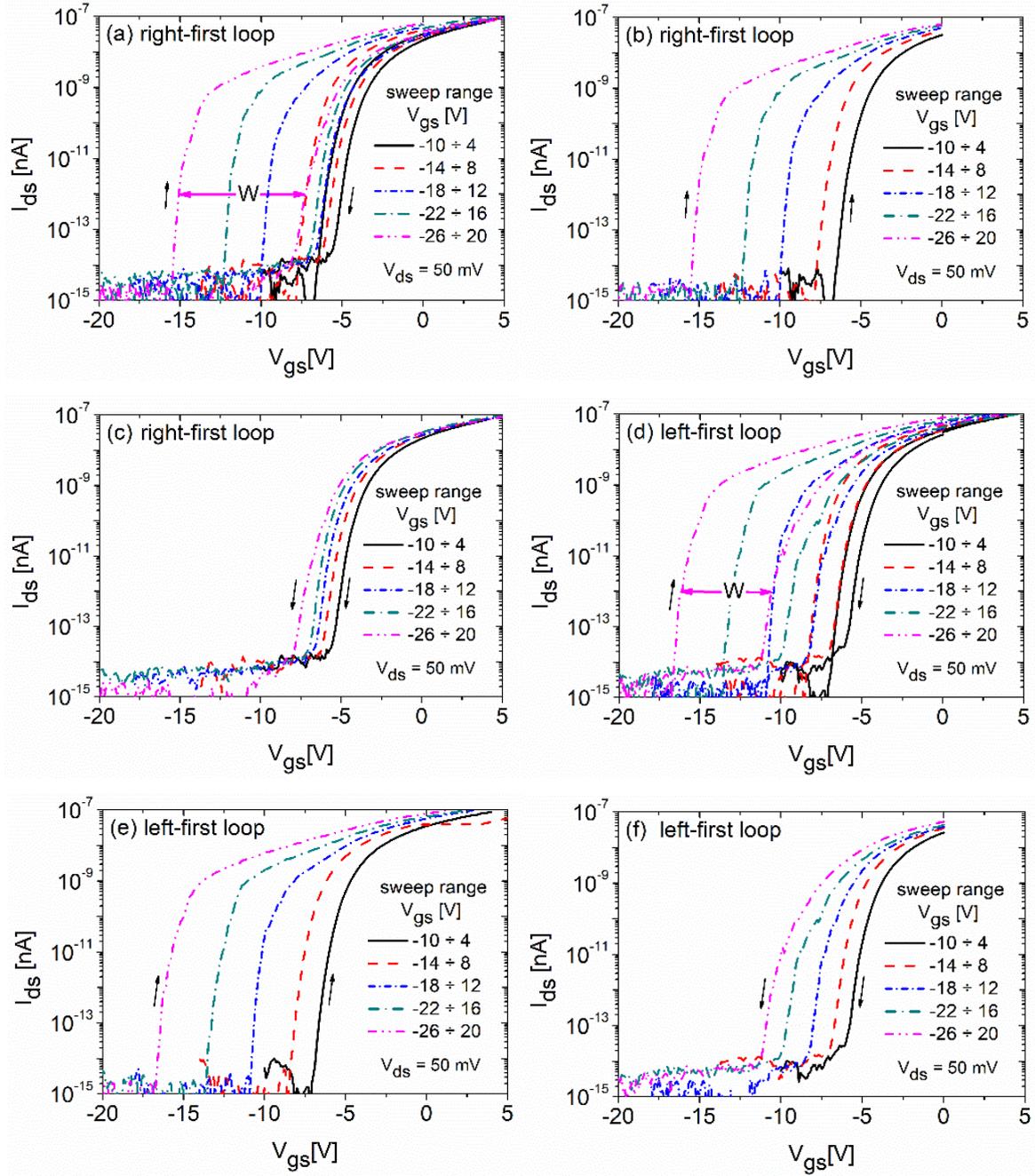

Figure 3. Transfer characteristics of the MoS$_2$ transistor for the back-gate voltage, $V_{gs}$, in loops of different amplitudes but with fixed steps ($\Delta V_{gs}$ =0.1 V). (a) Right-first loop: $V_{gs}$ increases from 0 to a positive voltage (e.g. 5 V), then is decreased to a negative voltage (e.g. -20 V) and finally increased again till 0 V. Branches of the transfer characteristic loops: (b) Rightward sweep (e.g. $V_{gs}$: -20 V → 0 V) and the (c) leftward sweep (e.g. $V_{gs}$: 5 V → -20 V). (d) Left-first loop: $V_{gs}$ decreases from 0 to a negative voltage (e.g. -20V), then is increased to a positive voltage (e.g. 5 V) and finally decreased again till 0 V. Branches of the transfer characteristic loops: (e) Rightward sweep (e.g. $V_{gs}$: -20 V → 5 V) and (f) leftward sweep ($V_{gs}$: 0 V → -20 V). All measurements are performed in dark, at room temperature and atmospheric pressure and with the drain bias $V_{ds} = 50\ mV$.

Their partial filling at $V_{gs} = 0\ V$ (Figure 4 (a)) accounts for the normally-on behavior of the $MoS_2$ transistor. As illustrated in Figure 4 (b) and (c), a positive gate voltage renders these states neutral, while a negative $V_{gs}$ make them charged with trapped positive charge. The trapping/detrapping process has a certain relaxation time, and the charge stored during the negative sweep can remain partially trapped during the positive sweep, thus causing a hysteresis. Residual $H_2O$ molecules at the $MoS_2/SiO_2$ interface, introduced during the fabrication process or adsorbed by the air-exposed device, can influence the charge transfer process by polarization. Figures 4 (d) and (e) show how water molecules can facilitate positive charge trapping and detrapping at a S vacancy by alternatively exposing $O^{2-}$ or $H^+$ ions.

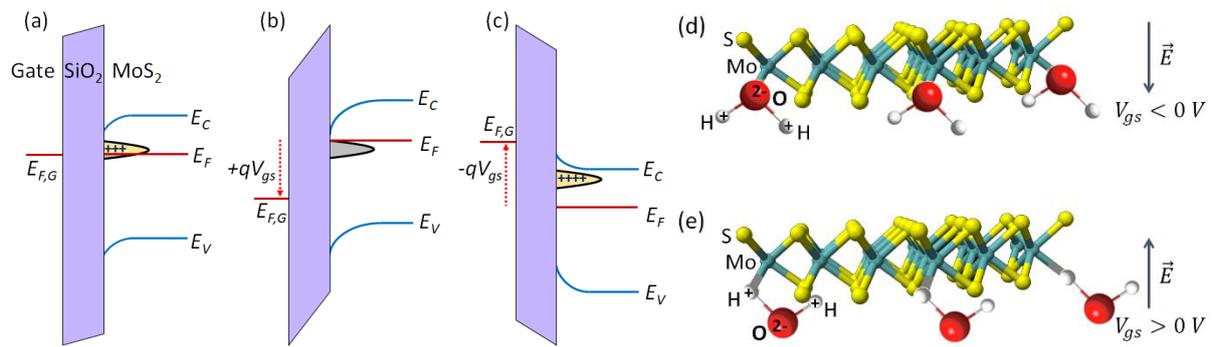

Figure 4. Band diagram of the $MoS_2/SiO_2/Si$ structure (Si is treated as degenerate), at zero (a), positive (b) and negative (c) gate bias, showing that a negative $V_{gs}$ favors the storage of positive charge at the $MoS_2/SiO_2$ interface. The distribution of donor-like traps has a peak at ~0.25-0.5 eV below the conduction band. Electron-filled states are neutral, while unoccupied states above the Fermi level (pictured in yellow color) are positively charged. Effect of negative (d) and positive (e) gate voltage on polarization of water molecules trapped at the $MoS_2/SiO_2$ interface.

To gain further insight in the hysteresis mechanism we studied it under different conditions. Figures 5 (a) and (b) show a linear dependence of the hysteresis width on $V_{gs}$ sweep range and time, which is expected due to the increased transfer/trapping time. Figures 5 (c) and (d) show that the hysteresis is dramatically affected by environmental conditions such as pressure and temperature, respectively. A decreasing pressure decreases the hysteresis and increases the mobility (see right-bottom inset of Figure 5(c)), confirming the important role of adsorbates and processing residues in enhancing the hysteresis. Pumping down till 15 mbar partially removes $O_2$ adsorbates and process contaminants such as carbon or resist from the exfoliation or fabrication process. Figure 5 (d) shows a steady growing hysteresis with temperature with a step up around room temperature. A smooth growth is observed also in the mobility vs. temperature plot (see inset), indicating that Coulomb scattering from trapped charges is dominant in the investigated temperature range. Remarkably, the same trend, but with increased hysteresis width, is present also when measurements are made under illumination (Figure 5 (e)).

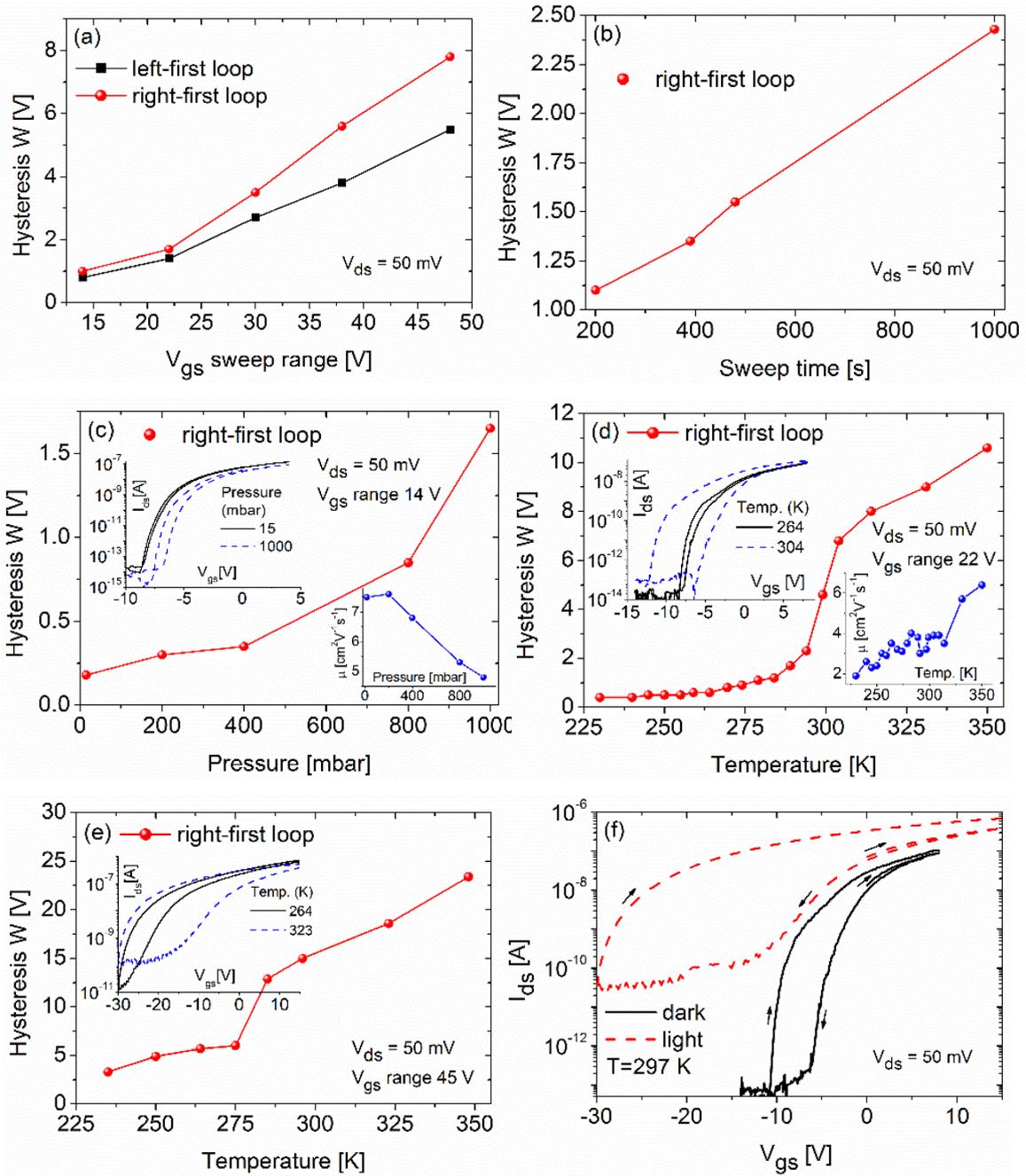

Figure 5. Dependence of the hysteresis width W on (a) $V_{gs}$ sweep range, (b) sweep time, (c) pressure (the top-left and bottom-right insets show two transfer curves at different pressures and the mobility vs pressure, respectively), (d) temperature in dark and (e) temperature under illumination (the insets show two representative transfer curves at different temperatures and the mobility versus temperature). (f) Comparison of transfer characteristics in dark and under $5 \ mW/cm^2$ illumination from a white LED system. All measurements are performed at drain bias $V_{ds} = 50 \ mV$.

Light substantially increases the channel conductance (Figure 5 (f)) for photoconductive and photogating effects, as we reported in a previous work [23]). Both temperature and light increase the availability of thermally or photo-generated minority carriers (holes) in $MoS_2$ and enable additional processes of positive charge trapping, which further enhance the hysteresis. We believe that the step observed in the temperature dependence is caused by $H_2O$ molecules. Water absorption on $MoS_2$ has been studied in connection to the degradation of $MoS_2$ lubrication properties in humid environments. It has been shown that, although $MoS_2$ is hydrophobic, edges and vacancies can be sensitive locations for water molecular adsorption, which has the maximum rate at ordinary room temperatures [66, 67].

The two current states at a given gate voltage due to hysteresis can be used as the two logic states of a solid-state memory. As a proof of concept, we show that the device can be switched between a high (On) and a low (Off) current state by applying a negative (erase) or positive (write) pulse, respectively as shown in Figure 6 (a) and (b). It is important noting the behavior of the low state current on the read cycle following the positive $V_{gs}$ pulse, in Figure 6 (a). The initial decreasing current (negative current transient) is a well-known in current-mode deep level transient spectroscopy (I-DLTS) as a signature of hole capture [68-70], which is another confirmation of the effective positive-charge trapping model of hysteresis proposed here. The exponential fittings in figure 6 (a) show hole trapping times with a typical time constant of 30 s. With time, the gating effect due to captured holes can take over and invert the transient behavior of the current.

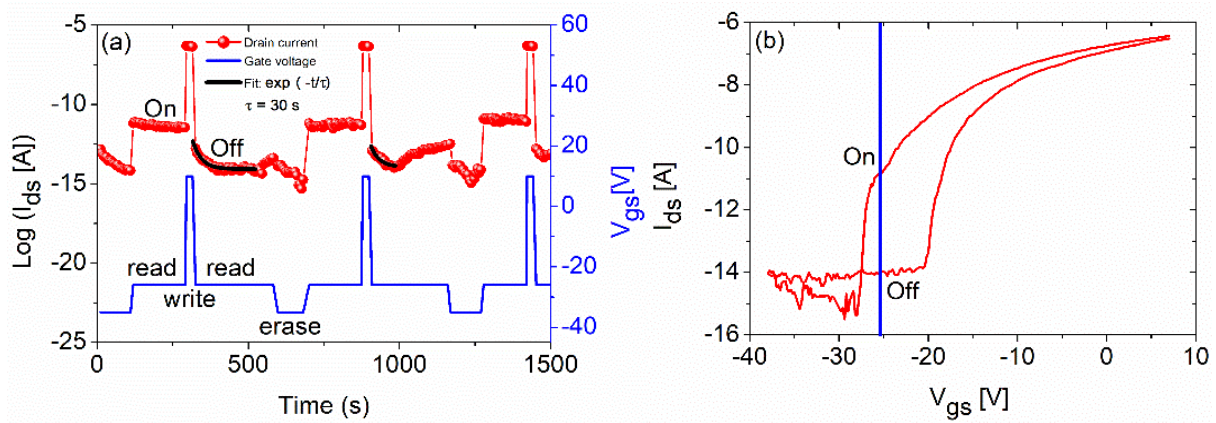

Figure 6. (a) On/Off current switching of the $MoS_2$ transistor used as memory device. Write and erase (set and reset) is achieved with $V_{gs}$ pulses of 10 V and -35 V, respectively, while $V_{gs} = -26\ V$ is the gate voltage for the reading operation. The black curves correspond to the fit of exponential decays $\sim e^{-t/\tau}$ from which the trapping time $\tau$ is evaluated. (b) Transfer characteristic of the device used in (a) with highlighted reading gate voltage.

The relatively long trapping/detrapping time constant opens the way to exploit pulse-mode measurements to obtain hysteresis-free transfer characteristics, which may be used for accurate estimation of device parameters like the electron mobility. Indeed, gate voltage pulses of 1 ms width and 100 ms periods have been shown to yield quasi hysteresis-free transfer characteristics [71].

As a final remark, we note that also the dependence of the hysteresis width on the illumination can be used for integral light intensity detection.

**Conclusions**

We studied hysteresis effects in back-gated $MoS_2$ transistors as a function of electrical stress and various environment conditions such as pressure, temperature and light. We concluded that a contribution to the hysteresis comes from S vacancies in the $MoS_2$ layer that behave as effective positive-charge trap centers. We pointed out that charge transfer from/to such centers is facilitated by the polarization of water molecules. We also showed that $O_2$ adsorption or exfoliation residues on $MoS_2$ unpassivated surface enhance the hysteresis. The same effect is caused by temperature and light, which both increase the availability of photo- or thermal-generated minority carriers and enable additional trapping processes. Finally, we pointed out that the hysteresis can be exploited in memory devices or in photodetectors.

**Acknowledgment**

We acknowledge the economic support of POR Campania FSE 2014-2020, Asse III Ob.Specifico l4, Avviso pubblico decreto dirigenziale n. 80 del 31/05/2016 and L.R. num. 5/2002 Finanziamento progetti annualità 2008, Prot. 2014, 0293185, 24/04/2014